\documentclass{ifacconf}

\usepackage{graphicx}
\usepackage{amssymb,amsmath}
\usepackage{epstopdf}
\renewcommand\phi{\varphi}
 \DeclareMathOperator{\col}{col}
 \DeclareMathOperator{\diag}{diag}
\newcommand{\trn}{^{\rm\scriptscriptstyle T}}
\DeclareMathOperator{\Real}{Re}
\usepackage{natbib}        % required for bibliography
\begin{document}
\begin{frontmatter}

\title{Synchronization in Networks with Nonlinearly Delayed Couplings on Example of Neural Mass Model} 

\thanks[footnoteinfo]{This work was performed in IPME RAS and supported by Russian Science Foundation (project no. 21-72-00107).}

\author[First]{Sergei A. Plotnikov} 

\address[First]{Institute for Problems of Mechanical Engineering, Saint-Petersburg, Russia, (e-mail: waterwalf@gmail.com)}

\begin{abstract}                % Abstract of not more than 250 words.
The problem of synchronization in heterogeneous networks of linear systems with nonlinear delayed diffusive coupling is considered. The network is presented in new coordinates mean-field dynamics and synchronization errors. Thus the problem of network synchronization is reduced to the studying of synchronization-error system stability. The circle criterion for time-delay systems is used to derive the stability conditions of synchronization-error system. Obtained results are applied to a network of neural mass model populations, and the synchronization conditions are established. Simulation results are provided to illustrate the obtained analytical results.
\end{abstract}


\end{frontmatter}
%===============================================================================

\section{Introduction}

Synchronization in networks of coupled oscillators is attractive phenomenon to study for specialists in various fields of science \cite{ARE08}. Synchronization underlies many natural phenomena and is the cornerstone of many technical concepts and engineering approaches. Examples of synchronization, include, among others, numerous forms of collective behavior in complex biological and artificial systems, such as flocks of birds, swarming, and rendezvous \cite{HER16,SUM10}; ensembles of oscillators \cite{HON11} and a group of mobile robots \cite{REN08}. Special attention is paid to synchronization in neural network dynamics. Synchronization depends on various network parameters, and a time delay in a signal propagation between the nodes plays a crucial role in this phenomenon.

Time delays are always present in real physical systems, therefore, in order to develop adequate realistic models of dynamic networks, one should to take into account delays in signal propagation in order to properly analyze the design of their dynamics. In neural networks, time delays can induce various rhythmic spatiotemporal patterns \cite{COO09,SON09}, change the stability of existing patterns \cite{ERM09}, and play a crucial role in synchronization behavior \cite{DAH09,SCH09}. Various works are devoted to study synchronization in delay coupled networks, just to mention a few, \cite{STE12,PRO13b,SEL15,PLO18}.

This paper continuous the work started in \cite{PLO21}. Here synchronization in heterogeneous networks of linear systems with nonlinear delayed diffusive coupling is considered. The network of neural mass model (NMM) populations \cite{JAN95} is an example of the networks of this type. The problem of network synchronization is reduced to studying the stability of synchronization error system which can be obtained using a coordinate transformation proposed in \cite{PAN17}. The analogue of the circle criterion for time-delay systems (TDS) proposed in \cite{CHU95} and generalized in \cite{BRY19} can be used to study the system stability. %The advantage of this approach over descriptor method and its variations proposed by E.~Fridman \cite{FRI03} is that one can obtain an analytical formula instead of linear matrix inequality. This fact is very important for studying the networks of large number of nodes because linear matrix inequalities of high dimensions need a lot of time for computation.

The rest of the paper is organized as follows. Section~\ref{sec:prel} reminds some important concepts related to synchronization and the circle criterion for TDS. In Sec.~\ref{sec:main} synchronization conditions of linear network with nonlinear delayed diffusive couplings are obtained, and NMM network is considered as an example Section~\ref{sec:sim} provides numerical results on synchronization. Finally, conclusions are given in Sec.~\ref{sec:conclusion}.

{\bf Notation.} Throughout the paper the superscript $\trn$ ($^*$) stands for matrix transposition (complex conjugate); $\mathbb{R}^n$ denotes the $n$ dimensional real Euclidean space with vector norm $|\cdot|$; $j=\sqrt{-1}$ is the imaginary unit; the notation $z=\col(x,y)$ means that $z$ is a vector of two components $x,~y$; the notation $D=\diag\{d_1,\dots,d_n\}$ means that $D$ is a $n\times n$ diagonal matrix, where $d_i$ is its $i$th diagonal element; $A\otimes B$ means the Kronecker product of matrices $A$ and $B$; $I_n$ is an identity $n\times n$ matrix, while $0_n$ is a $n\times n$ matrix of zeros.

\section{Preliminaries}\label{sec:prel}
\subsection{Circle Criterion for Time-Delay Systems}\label{sec:circle}
In this section the details about studying the stability of linear system with multiple nonlinearities with time-varying delays are given. The analogue of the circle criterion proposed in \cite{CHU95} and its generalization for multi input -- multi output (MIMO) systems proposed in \cite{BRY19} will be used with the purpose of studying the global stability of nonlinear delay-coupled networks.

Consider the system, which is described by the following equations
\begin{equation}\label{c1}
\begin{aligned}
\dot x&(t)=Ax(t)+B\phi(t,\sigma(t-\tau(t))), \\
\sigma&(t-\tau(t))=C\trn x(t-\tau(t)),
\end{aligned}
\end{equation}
where $A$ is a constant $n\times n$ matrix, $B$, $C$ are constant $n \times m$ matrices and $x\in\mathbb{R}^n$ is a state vector, \mbox{$\sigma =\col\{\sigma_1,\dots,\sigma_m\}\in\mathbb{R}^m$} is an input of the system, \mbox{$\tau_i=\col\{\tau_1,\dots,\tau_m\}\in\mathbb{R}^m$}, $\tau_i(t)\in[0,T]$, $i=1,\dots,m$, $\forall t$ is a bounded time-varying delay, $\phi=\col\{\phi_1,\dots,\phi_m\}$ is a vector function of sector-bounded nonlinearities, i.e. the following inequalities are fulfilled:
\begin{equation}\label{c2}
\begin{gathered}
\mu_{11}\le\phi_1(t,\sigma_{1})/\sigma_{1}\le \mu_{21}, \\
\vdots \\
\mu_{1m}\le\phi_m(t,\sigma_{m})/\sigma_{m}\le \mu_{2m}, 
\end{gathered}
\end{equation}
for $\sigma_i\ne0$, $i=1,\dots,m$.
\begin{thm}
\noindent For the system \eqref{c1} denote the transfer function of its linear part $W(p)=C\trn(A-p I_n)^{-1}B$ and the characteristic polynomial \mbox{$\Delta(p)=\det(p I_n-A)$}. Introduce the following diagonal matrices $\mu_1=\diag\{\mu_{11},\dots,\mu_{1m}\}$, $\mu_2=\diag\{\mu_{21},\dots \mu_{2m}\}$.
Let the following assumptions be fulfilled:
\begin{enumerate}
\item nonlinearity $\phi(\sigma)$ in system \eqref{c1} satisfies the inequalities \eqref{c2} for $\sigma_i\ne0$, \mbox{$i=1,\dots,m$};
\item There exists diagonal $m\times m$ matrix $\mu_0$ such that each element $\mu_{0i}$ lies between $\mu_{1i}$ and $\mu_{2i}$, $i=1,\dots,m$, and matrix $A+B\mu_0 C\trn$ is Hurwitz;
\item For some diagonal $m \times m $ matrix $\nu$ with positive diagonal elements $\nu_i$ such that \mbox{$1-4\nu_i\mu_{1i}\mu_{2i}>0$}, \mbox{$i=1,\dots,m$} the function
\begin{multline}\notag
\pi(\omega)=W(j\omega)^*(\mu_1\mu_2\nu-T^2\omega^2I_m/4)(I_m-4\nu\mu_1\mu_2)\\
\times W(j\omega)+\Real[W(j\omega)(\mu_1+\mu_2)(I_m-4\nu\mu_1\mu_2)\nu]\\
+\nu(I_m-\nu(\mu_1+\mu_2)^2),
\end{multline}
satisfies the following conditions
\begin{subequations}
\begin{eqnarray}\label{c4a}
\lim\limits_{\omega\to\infty}\pi(\omega)&>&0,\\
|\Delta(j\omega)|^2\pi(\omega)&>&0, \quad \forall\omega\ge 0, \label{c4b}
\end{eqnarray}
\end{subequations}
(for $\omega$ such that $|\Delta(j\omega)|=0$ holds the inequality \eqref{c4b} is understood as a limiting one).
\end{enumerate}
Then there exist positive constants $C_1$, $C_2$, $\epsilon$ depending only on the coefficients of the linear part of the system \eqref{c1} $A$, $B$, $C$ and matrices $\mu_1$, $\mu_2$, $TI_n$, such that for all solutions $x(t)$ of the system \eqref{c1} with a continuous initial function $x_0(t)$ defined for $t\in[-\tau_{\max},0]$, where $\tau_{\max}=\max_{i=1,\dots,m}\tau_i(0)$, the following inequality holds
$$
\|x(t)\|\le(C_1\|x_0(0)\|+C_2\max\limits_{-\tau_{\max}\le t\le 0}\|C\trn x_0(t)\|)\e^{-\epsilon t},
$$
for all $t\ge0$.
\end{thm}

\subsection{Synchronization}\label{sec:synch}
Here the mathematical notion of synchronization will be introduced. The networks of linear systems with heterogeneous delayed nonlinearities are considered throughout this paper. Each system of this type can be presented in normal form as the following:
\begin{equation}\label{m1}
\begin{aligned}
\dot y_i(t)&=A^yx_i(t)+\phi_i(\sigma_i(t-\tau_i(t))), \\
\dot z_i(t)&=A^zx_i(t), \quad i=1,\dots,N,\\
\end{aligned}
\end{equation}
where $y_i\in\mathbb{R}^m$, $z_i\in\mathbb{R}^{n-m}$, $x_i=\col\{y_i,z_i\}\in\mathbb{R}^n$ and $\sigma_i\in\mathbb{R}^m$ denote the input, the zero-dynamics, the state and the output of the $i$th system, respectively. $A^y\in\mathbb{R}^{m\times n}$ and $A^z\in\mathbb{R}^{(n-m)\times n}$ are constant matrices. $\phi_i=\col\{\phi_{i1}\dots\phi_{im}\}:\mathbb{R}^m\to\mathbb{R}^m$ is a vector function, while $\tau_i=\col\{\tau_{i1},\dots,\tau_{iN}\}$ is a time-varying delay. As input diffusive coupling is considered, which is described by
\begin{equation}\label{contr}
\sigma_i(t-\tau_i(t))=\sum\limits_{k=1}^N \gamma_{ik}[y_i(t-\tau_{ik}(t))-y_k(t-\tau_{ik}(t))].
\end{equation}
Suppose that the graph of the network under consideration is connected and undirected, therefore $\gamma_{ik}=\gamma_{ki}$ \mbox{$\forall i\ne k$}, $i,k=1,\dots,N$. For this type of coupling both signals are time-delayed. Such type of coupling may be observed, for instance, when the systems are interconnected by a centralized control law.

Synchronization phenomenon is often defined as the asymptotically identical evolution
of the systems. One can easily introduce the notion of the asymptotic coordinate synchronization \cite{FRA07}:
\begin{equation}\label{goal1}
\lim\limits_{t\to\infty}(x_i(t)-x_k(t))=0,\quad i,k=1,\dots,N.
\end{equation}
The fulfilment of \eqref{goal1} means that the asymptotic behavior of all nodes of the network \eqref{m1}, \eqref{contr} is identical and can be described by the function $x_s\in\mathbb{R}^n$. Thus by defining the synchronization errors as $e_i=x_i-x_s$ one can study the problem of network synchronization as the problem of stability of synchronization-error system (SES).

To obtain the equations of SES the approach proposed in \cite{PAN17} can be used. The idea behind this approach is the following: the system state space is decomposed in two orthogonal subspaces, one on which is projected the behavior of the mean-field state and one in which lay the synchronization errors. To obtain the network equations in new coordinates suppose that the graph of considered network is connected and undirected. Following the steps described in \cite{PLO21} up to a time-delay one can present the equations of the network \eqref{m1}, \eqref{contr} in the following form:
\begin{subequations}
\begin{eqnarray}\notag
\dot{\tilde x}_1(t)&=& A\tilde x_1(t) + (\mathbf{1}_N\otimes E_m)\\\label{mf}
&\times&\Phi [(LU_1\otimes E_m\trn)\tilde x_2(t-\tau(t))],\\\notag
\dot{\tilde x}_2(t)&=& (I_{N-1}\otimes A)\tilde x_2(t) + (U_1^\dagger\otimes E_m)\\
&\times&\Phi [(LU_1\otimes E_m\trn)\tilde x_2(t-\tau(t))],\label{err}
\end{eqnarray}
\end{subequations}
where 
\begin{equation}\label{main}
\begin{aligned}
A=\begin{bmatrix}A^y \\ A^z\end{bmatrix}\in\mathbb{R}^{n\times n},\quad \tau(t)=\begin{bmatrix}\tau_1(t) \\ \vdots \\ \tau_N(t),\end{bmatrix}\in\mathbb{R}^{N^2}\\
\Phi(\sigma)=\begin{bmatrix}\phi_1(\sigma_1) \\ \vdots \\ \phi_N(\sigma_N) \end{bmatrix}: \mathbb{R}^{mN}\to\mathbb{R}^{mN}
\end{aligned}
\end{equation}
are the matrix of the linear part of the individual system, the vector of delays and the nonlinear vector function, respectively;
\begin{equation}\label{lapl}
L=\begin{bmatrix} 
\sum_{k=2}^{N}\gamma_{1k} &-\gamma_{12} &\cdots&-\gamma_{1N}\\
-\gamma_{21} &\sum_{k=1,k\ne2}^{N}\gamma_{2k}&\cdots&-\gamma_{2N} \\
\vdots &\vdots &\ddots&\vdots \\
-\gamma_{N1} &-\gamma_{N2} &\cdots&\sum_{k=1}^{N-1}\gamma_{Nk}
\end{bmatrix}\in\mathbb{R}^{N\times N},
\end{equation}
is the Laplace matrix defining the coupling links in the network, where $\gamma_{ik}=\gamma_{ki}$ $\forall i\ne k,~i,k=1,\dots,N$ by the assumption;
\begin{equation}\label{supp}
\begin{aligned}
\mathbf{1}_N=\begin{bmatrix}1 \\ \vdots \\1\end{bmatrix}\in\mathbb{R}^N, \quad E_m=\begin{bmatrix}I_m \\0_{(m-n)\times n}\end{bmatrix}\in\mathbb{R}^{n\times m}, \\
U=\begin{bmatrix}\mathbf{1}_N & U_1\end{bmatrix}\in\mathbb{R}^{N\times N}, \quad U^{-1}=\begin{bmatrix}\mathbf{1}_N\trn \\ U^\dagger_1 \end{bmatrix}\in\mathbb{R}^{N\times N} \\
\mbox{such that~}U^{-1}LU=\diag\{\lambda_1,\dots,\lambda_N\} \\
\end{aligned}    
\end{equation}
are the auxiliary matrices and vectors;
$$
x=\begin{bmatrix}x_1 \\ \vdots \\x_N\end{bmatrix}\in\mathbb{R}^{nN}, ~{\tilde x}_1=\mathbf{1}_N\trn\otimes I_n x, ~{\tilde x}_2= U_1^\dagger\otimes I_n x
$$
 are the state space vector of the whole network, the vector proportional to the mean-field dynamics (MFD), and the vector proportional to the synchronization error.

\subsection{Neural Mass Model}
As an example of linear system with nonlinear delayed couplings a time-delay NMM will be considered. This model is based on the standard NMM proposed in \cite{JAN95} and incorporates a time delay \cite{GEN14}. The presence of a time delay in neuronal signal transmission could cause seizure-like activity in the brain. The NMM simulates the average firing activity of a population of pyramidal neurons that interacts with two populations of intercalary neurons and integrates inhibitory and excitatory signals from them \cite{JAN95,GRI06}. The dynamics of each neuronal population is described by two first-order nonlinear delay differential equations as follows:
$$
\begin{aligned}
\dot y_1(t) &= -2\alpha y_1(t) -\alpha ^2y_2(t) + \alpha\beta\phi[\sigma(t-\tau(t))], \\
\dot y_2(t) &= y_1(t).
\end{aligned}
$$
where $y_2$ refers to the post-synaptic potential, i.e. the deviation of the membrane from the resting potential, while $y_1$ is its derivative, $\sigma$ is a delayed input. The parameters $\alpha$ and $\beta$ are different in the excitatory and inhibitory cases. $\alpha$ is the reciprocal of the synaptic/membrane time constant; $\beta$ is the gain for the post-synaptic response kernel; $\tau(t)$ is a time-varying delay. The function $\phi(\sigma)$ is a nonlinear centered sigmoidal function relating the neuronal states
\begin{equation}\label{sigm}
\phi(\sigma) = \frac{g}{1+\e^{k_0-\sigma}}-\frac{g}{1+\e^{k_0}},
\end{equation}
where $k_0$ represents the ratio of average inhibitory synaptic gain and $g>0$ is the average excitatory synaptic gain.

\section{Main Result}\label{sec:main}
\subsection{Synchronization Conditions. General Case}
This paper considers a network of $N$ linear dynamical systems with heterogeneous delayed nonlinear couplings in normal form \eqref{m1}. The graph of considered network is supposed to be connected and undirected, and the connections between the nodes of the network are diffusive ones \eqref{contr}. As described in the Subsec.~\ref{sec:synch}, such a network can be represented in the form of MFD \eqref{mf} and SES \eqref{err}. To study the synchronization problem of such network one can apply the circle criterion for TDS (see Subsec.~\ref{sec:circle}) to the SES \eqref{err} to obtain the conditions of its stability. These conditions in turn will guarantee synchronization in the original network.

The transfer function of the linear part of the system \eqref{err} can be calculated just like in \cite{PLO21}:
\begin{equation}\label{tf}
W(p)=L\otimes[ E_m\trn (A-pI_{n})^{-1}E_m],
\end{equation}
where $L$ is the Laplace matrix \eqref{lapl}, $A$ is the matrix of linear part \eqref{main} and $E_m$ is the supplementary matrix \eqref{supp}.

The nonlinear part of the system \eqref{err} are described by the functions $\phi_i=\col\{\phi_{i1}\dots\phi_{im}\}$, $i=1,\dots, N$. Suppose that they belong to the two-cavity sector between two straight lines, i.e. the following inequalities are fulfilled:
\begin{equation}\label{bounds}
\begin{gathered}
\mu_{1i1}\le\phi_{i1}(\sigma_{i1})/\sigma_{i1}\le \mu_{2i1}, \\
\vdots \\
\mu_{1im}\le\phi_{im}(\sigma_{im})/\sigma_{im}\le \mu_{2im}, \\
i=1,\dots, N.
\end{gathered}
\end{equation}
Introduce diagonal matrices matrices by the following way 
$$
\begin{gathered}
\mu_1=\diag\{\mu_{111},\dots,\mu_{1m1},\dots,\mu_{11N},\dots,\mu_{1mN}\}, \\
\mu_2=\diag\{\mu_{211},\dots,\mu_{2m1},\dots,\mu_{21N},\dots,\mu_{2mN}\}.
\end{gathered}
$$ 
Suppose that all delays $\tau_i=\col\{\tau_{i1},\dots,\tau_{iN}\}$, $i=1,\dots,N$ in system \eqref{err} are bounded functions, i.e. $\tau_i(t)\in[0,T]$, $i=1,\dots,m$, $\forall t$.

Thus the theorem about synchronization of linear networks with nonlinear delayed couplings can be formulated.
\begin{thm}\label{th2}
    If the following conditions are fulfilled 
\begin{enumerate}
    \item The network \eqref{m1}, \eqref{contr} nonlinearities lie in the sector, i.e. inequalities \eqref{bounds} hold; 
    \item The graph of the network \eqref{m1}, \eqref{contr} is connected and undirected.
    \item There exists matrix $\mu_0\in\mathbb{R}^{mN}\times\mathbb{R}^{mN}$ such that matrices $\mu_0-\mu_1$ and $\mu_2-\mu_0$ have only nonnegative elements, and matrix 
$$
    \Psi=I_{N-1}\otimes A+[U_1^\dagger\otimes E_m]\mu_0 [LU_1\otimes E_m\trn]
$$
    is Hurwitz;
    \item 
    For some diagonal $mN \times mN$ matrix $\nu$ with positive diagonal elements such that matrix $I_{mN}-4\nu\mu_{1}\mu_{2}$ has positive diagonal elements, the function
\begin{multline}\notag
\pi(\omega)=W(j\omega)^*(\mu_1\mu_2\nu-T^2\omega^2I_{mN}/4)\\
\times(I_{mN}-4\nu\mu_1\mu_2)W(j\omega)\\
+\Real[W(j\omega)(\mu_1+\mu_2)(I_{mN}-4\nu\mu_1\mu_2)\nu]\\
+\nu(I_{mN}-\nu(\mu_1+\mu_2)^2),
\end{multline}
satisfies the following conditions
\begin{subequations}
\begin{eqnarray}\label{m4a}
\lim\limits_{\omega\to\infty}\pi(\omega)&>&0,\\
|\Delta(j\omega)|^2\pi(\omega)&>&0, \quad \forall\omega\ge 0, \label{m4b}
\end{eqnarray}
\end{subequations}
(for $\omega$ such that $|\Delta(j\omega)|=0$ holds the inequality \eqref{m4b} is understood as a limiting one), where $W(p)$ is the transfer function \eqref{tf} and 
\begin{multline}\notag
\Delta(p)=\det(pI_{n(N-1)}-I_{N-1}\otimes A)\\
=\det(I_{N-1}\otimes(pI_n-A))=(\det(pI_n-A))^{N-1}
\end{multline}
is the characteristic polynomial of the matrix of the linear part \eqref{err}.
\end{enumerate}
Then the systems in the network \eqref{m1}, \eqref{contr} are asymptotically synchronized.
\end{thm}

\subsection{Synchronization Conditions of Neural Mass Model Populations}

This section considers the heterogeneous network of nonlinearly delayed coupled NMM populations
\begin{equation} \label{net}
\begin{aligned}
\dot y_{1i}(t) &= -2\alpha y_{1i}(t) -\alpha ^2y_{2i}(t) \\
&+ \alpha\beta\phi_i\left\{\sum\limits_{k=1}^N \gamma_{ik}[y_{1i}(t-\tau_{ik}(t))-y_{1k}(t-\tau_{ik}(t))]\right\}, \\
\dot y_{2i}(t) &= y_{1i}(t), \quad i=1,\dots,N,
\end{aligned}
\end{equation}
where $x=\col\{y_{11},y_{21},\dots,y_{1N},y_{2N}\}$ is a state vector; $\alpha$, $\beta$ are system parameters; $\gamma_{ik}$ are coupling coefficients; $\tau_{ik}(t)\in[0,T]$, $\forall t$, $i,k=1,\dots,N$ are bounded time-varying delays (all delay functions have the same upper bound $T$). Functions $\phi_i$ are sigmoidal ones, which are descirbed by \eqref{sigm} with parameters $g_i$ and $k_{0i}$, $i=1,\dots,N$.

The network of NMMs can be presented in new coordinates \eqref{mf}, \eqref{err} with matrices
$$
A=\begin{bmatrix}-2\alpha & -\alpha ^2 \\1 & 0\end{bmatrix},\quad E_m=\begin{bmatrix} 1 \\ 0\end{bmatrix}.
$$

To study the network \eqref{net} synchronization one should check the conditions of Theorem~\ref{th2}:

Sigmoidal functions \eqref{sigm} are sector ones, which lie between two straight lines $0$ and $0.5g_i\sigma$ (see explanation in \cite{GOR17}), $i=1,\dots,N$. One can find $g_{\max}=\max_{i=1,\dots,N}g_i$. Suppose that $\alpha\beta>0$, then the matrices from the condition (1) of the Theorem~\ref{th2} can be expressed as $\mu_1=0_{N}$ and $\mu_2=0.25g_{\max}\alpha\beta I_{N}$. 

Supposing that the graph of considered network \eqref{net} is connected and undirected guarantees the fulfillment of the condition (2) of the Theorem~\ref{th2}.

The condition (3) of Theorem~\ref{th2} is the same as for the case without delays: this fact follows from the conditions of Circle criterion. This condition was previously checked in \cite{PLO21}: If $\alpha>0$, then the matrix $\Psi$ in the condition (3) of the Theorem~\ref{th2} is Hurwitz.

To calculate the function $\pi(\omega)$ from the condition (4) of the Theorem~\ref{th2} the frequency transfer function $W(j\omega)$ should be found using formula \eqref{tf}:
\begin{multline}\label{tfex}
W(j\omega)=L\otimes\left[\begin{bmatrix} 1 & 0\end{bmatrix} \begin{bmatrix}-2\alpha-j\omega & -\alpha ^2 \\1 & -j\omega\end{bmatrix}^{-1} \begin{bmatrix} 1 \\ 0\end{bmatrix}\right]\\=
\frac{-j\omega }{\alpha^2-\omega^2+2j\alpha\omega }L=\frac{-2\alpha\omega^2-j\omega(\alpha^2-\omega^2) }{(\alpha^2+\omega^2)^2}L.
\end{multline}

Meaning that $\mu_1=0_{N}$ and $\mu_2=0.25g_{\max}\alpha\beta I_{N}$ and choosing matrix $\nu$ as $\nu_0 I_{N}$ one obtains:
\begin{multline}\notag
\pi(\omega)=\frac{-T^2\omega^2}{4}W(j\omega)^*W(j\omega)+\frac{g_{\max}\alpha\beta\nu_0}{4}\Real[W(j\omega)]\\
+\nu_0\left(1-\frac{\nu_0g_{\max}^2\alpha^2\beta^2}{16}\right)I_{N}.
\end{multline}

Since the graph of considered network is undirected the corresponding Laplace matrix $L$ is symmetric. Using this fact and \eqref{tfex} the function $\pi(\omega)$ equals:
\begin{multline}\notag
\pi(\omega)=\frac{-T^2\omega^4}{4(\alpha^2+\omega^2)^2}L^2-\frac{g_{\max}\alpha^2\beta\nu_0\omega^2}{2(\alpha^2+\omega^2)^2}L\\
+\nu_0\left(1-\frac{\nu_0g_{\max}^2\alpha^2\beta^2}{16}\right)I_{N}.
\end{multline}

Now check the matrix inequality \eqref{m4a}:
$$
\lim\limits_{\omega\to\infty}\pi(\omega)=\frac{-T^2}{4}L^2+\nu_0\left(1-\frac{\nu_0g_{\max}^2\alpha^2\beta^2}{16}\right)I_{N}>0
$$
Consider some symmetric matrix $P$: it is positive definite if the corresponding quadratic form $x\trn Px$ is positive $\forall x\ne 0$. $\exists$ matrix $U$: $P=UDU\trn$, where $D$ is a diagonal matrix. Then for $z=U\trn x$ one obtains $z\trn Dz>0$. Therefore the obtained inequality is fulfilled if and only if the following inequalities are fulfilled:
\begin{equation}\label{ref}
\frac{-T^2\lambda_i^2}{4}+\nu_0\left(1-\frac{\nu_0g_{\max}^2\alpha^2\beta^2}{16}\right)>0, ~ i=1,\dots,N,
\end{equation}
where $\lambda_i$ are the eigenvalues of Laplace matrix $L$. Then the maximal value of the delay $T$ can be estimated:
\begin{equation}\label{eqs}
T^2<\frac{16\nu_0-\nu_0^2g_{\max}^2\alpha^2\beta^2}{4\lambda_i^2}, \quad i=1,\dots,N.
\end{equation}
There is a quadratic equation depending on $\nu_0$ in the numerator of resulting fraction, which has the maximal value for $\nu_0=8/(g_{\max}^2\alpha^2\beta^2)$. All equations \eqref{eqs} hold if
\begin{equation}\label{result}
    T<\frac{4}{g_{\max}\lambda_{\max}\alpha\beta},
\end{equation}
where $\lambda_{\max}$ is the maximal eigenvalue of Laplace matrix $L$.

From \eqref{m4b} one obtains that inequality
\begin{multline}\notag
\frac{-T^2\omega^4}{4}L^2-\frac{g_{\max}\alpha^2\beta\nu_0\omega^2}{2}L\\
+\nu_0\left(1-\frac{\nu_0g_{\max}^2\alpha^2\beta^2}{16}\right)(\alpha^2+\omega^2)^2I_{N}>0
\end{multline}
should be fulfilled $\forall\omega\ge 0$. As before the set of the following inequalities can be consider instead of the obtained matrix inequality:
\begin{multline}\label{help}
\left[\nu_0\left(1-\frac{\nu_0g_{\max}^2\alpha^2\beta^2}{16}\right)-\frac{T^2\lambda_i^2}{4}\right]\omega^4\\
+\alpha^2\nu_0\left[2\left(1-\frac{\nu_0g_{\max}^2\alpha^2\beta^2}{16}\right)-\frac{g_{\max}\beta\lambda_i}{2}\right]\omega^2\\
+\alpha^4\nu_0\left(1-\frac{\nu_0g_{\max}^2\alpha^2\beta^2}{16}\right)>0.
\end{multline}
The coefficient before $\omega^4$ is the same as \eqref{ref}, and it is positive if $\nu_0=8/(g_{\max}^2\alpha^2\beta^2)$ and the inequality \eqref{result} is fulfilled. Zero order term is also positive for chosen value of $\nu_0$. The coefficient before $\omega^2$ is positive if $g_{\max}\beta\lambda_{\max}<2$. Thus the inequality \eqref{help} is fulfilled for chosen parameters.

All conditions of the Theorem~\ref{th2} are fulfilled, therefore the network \eqref{net} is synchronized. The following theorem holds.
\begin{thm}\label{th3}
If the network \eqref{net} systems parameters $\alpha>0$ and $\beta>0$, $g_i>0$, $i=1,\dots,N$, the graph of the network is connected and undirected, the maximum eigenvalue of the Laplace matrix $L$ is less than $2/(\beta g_{\max})$, and the delays in the signal propagation are bounded \eqref{result}, then the network of NMMs is asymptotically synchronized.
\end{thm}
Note that while $T\to0$ and choosing $\nu_0\to0$ such that $T^2/\nu_0\to 0$, we get the similar conditions of network synchronization as in \cite{PLO21}.

\section{Simulation}\label{sec:sim}
This section presents the results of simulation. The network of NMM populations with $N=10$ is considered. The system parameters $\alpha$ and $\beta$ are equal to $1$ and $0.8$, respectively. The parameters defining the shape of sigmoidal function have uniform distribution: $g_i$, $i=1,\dots,N$ are distributed on the interval $[0;1]$, while $k_{0i}$, $i=1,\dots,N$ are distributed on the interval $[-1;1]$. The graph of considered network is weighted, connected and undirected, meaning that its adjacency matrix is a symmetric sparse matrix with density $0.7$, which means that it has approximately $0.7N^2$ nonzero entries. Let the delays $\tau_{ik}(t)\in[0;T]$, $i,k=1,\dots,N$ be time-varying functions
$$
\tau_{ik}(t)=h_{1ik}+h_{2ik}\sin(h_{3ik}t+h_{4ik}),
$$ 
which are uniformly distributed on the interval $[0;T/2]$ such that $h_{1ik}>h_{2ik}$, $i,k=1,\dots,N$. The maximum value of the delay $T$ will be defined later. The initial functions $y_{1i}(t)$, $y_{2i}(t)$, $t\in[-T;0]$, $i=1,\dots,N$ are constants, which are uniformly distributed on the interval $[-1;1]$.

\begin{figure}
\center{\includegraphics[width=1\linewidth]{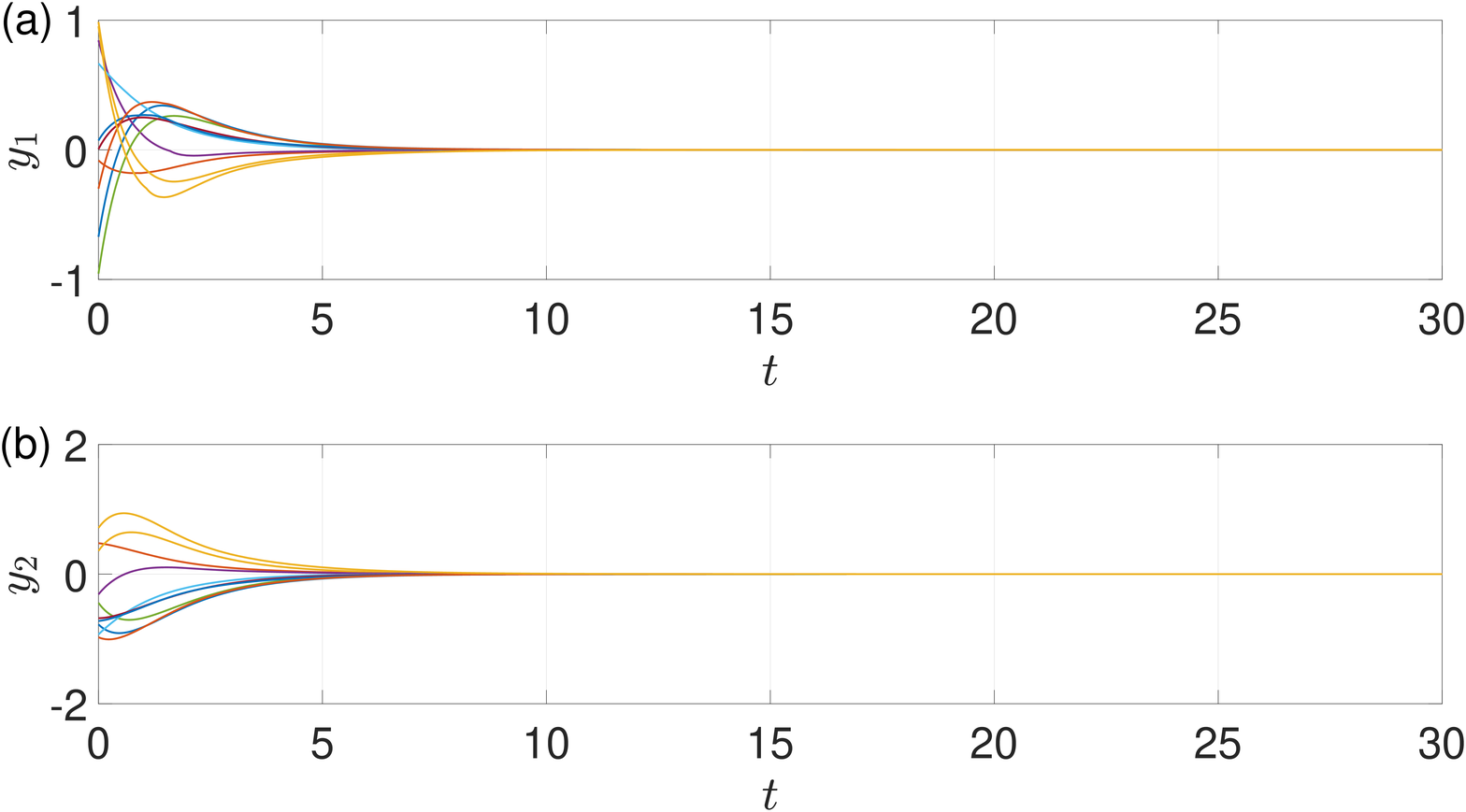}}
\caption{Synchronization of neural mass model (NMM) population network \eqref{net} with $N=10$ nodes. (a) and (b) dynamics of post-synaptic potential derivatives $y_1$ and post-synaptic potentials $y_2$ of all nodes, respectively. System parameters: $N=10$, $\alpha=1$, $\beta=0.8$, $T=2$, $g_i$, $i=1,\dots,N$ are uniformly distributed on the interval $[0;1]$, $k_{0i}$, $i=1,\dots,N$ are uniformly distributed on the interval $[-1;1]$, $\lambda_{\max}=2.5080$. Initial functions $y_{1i}(t)$, $y_{2i}(t)$, $t\in[-T;0]$, $i=1,\dots,N$ are constants, which are uniformly distributed on the interval $[-1;1]$.}
\label{fig1}
\end{figure}

First of all consider the case, when NMM network has parameters satisfying the conditions of the Theorem~\ref{th3}. The maximum eigenvalue $\lambda_{\max}$ of the Laplace matrix $L$ is equal to $2.5080$ is this case, while \mbox{$g_{\max}=0.9724$}. For these parameters of the network the inequality \mbox{$\lambda_{\max}<2/(\beta g_{\max})\approx 2.5709$} is fulfilled. Choosing $T=2$ one can ensure the fulfillment of the inequality \eqref{result}, thereby guarantee the fulfillment of all condition of the Theorem~\ref{th3}. This means that for these parameters the network of NMMs will synchronize. Figure~\ref{fig1} presents the results of simulation. As one can see, for the chosen parameter values, there is synchronization among the state variables of the network, and all system trajectories tend to equilibrium point, which confirms~Theorem~\ref{th3}.

\begin{figure}
\center{\includegraphics[width=1\linewidth]{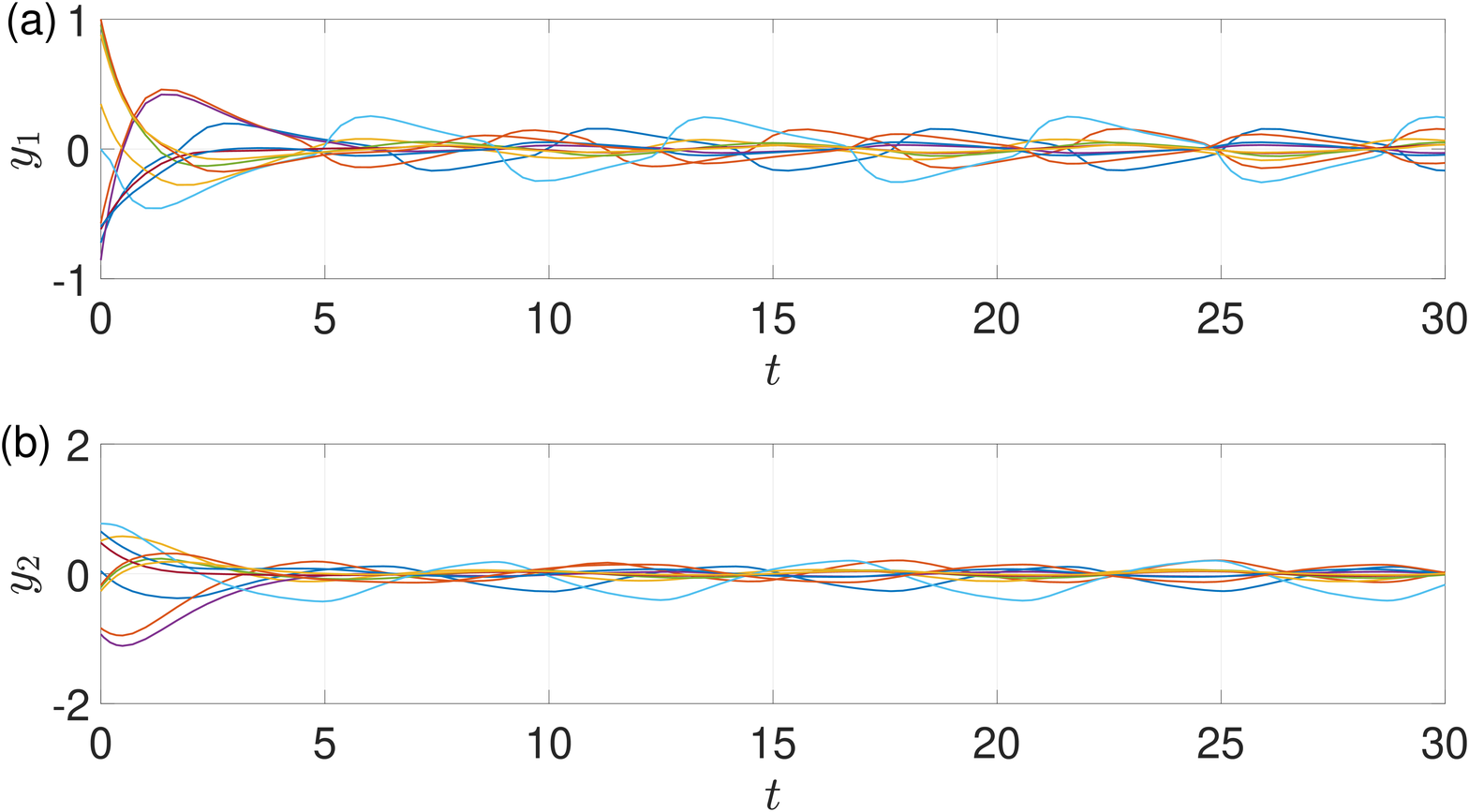}}
\caption{Desynchronization of neural mass model (NMM) population network \eqref{net} with $N=10$ nodes. (a) and (b) dynamics of post-synaptic potential derivatives $y_1$ and post-synaptic potentials $y_2$ of all nodes, respectively. System parameters: $T=10$, $\lambda_{\max}=1.7983$. Other parameters and initial functions are the same as in Fig.~\ref{fig1}.}
\label{fig2}
\end{figure}

Now consider the case, when the delays in signal propagation between the nodes are too large, that prevents the network synchronization. In this case $\lambda_{\max}=1.7983$ and $g_{\max}=0.9630$, which means that the inequality \mbox{$\lambda_{\max}<2/(\beta g_{\max})\approx 2.5960$} is also fulfilled. Choosing \mbox{$T=10$} one violates the Theorem~\ref{th3} condition \eqref{result}. Therefore for these parameters of the network, the Theorem~\ref{th3} doesn't guarantee the network synchronization. One can see the results of simulation in Fig~\ref{fig2}: there is no synchronization among the state variables of the network.

\section{Conclusion}\label{sec:conclusion}

In this paper, the problem of heterogeneous network synchronization of linear systems with delayed nonlinear diffusive couplings. The delays are supposed to be bounded time-varying functions. The nonlinear coupling functions can be different but their graphs should lie in two-cavity sector between two straight lines. As in \cite{PLO21} the synchronization problem is reduced to study the stability of the SES. The coordinate transformation approach proposed in \cite{PAN17} is used to present the network in coordinates "MFD - SES". To find the conditions for SES stability the circle criterion for TDS was applied. The theorem about network synchronization of this type was formulated and proven. 

As an example, the dynamics of NMM populations connected via delayed nonlinear diffusive coupling was considered. Using the obtained theorem, the simple condition for network synchronization was derived. In the case of the delay absence this condition coincide with the result obtained in \cite{PLO21}. Also, the simulation of NMM network dynamics was performed. In the case, when the Theorem~\ref{th3} conditions are fulfilled, one can observe the synchronization between the network states. In the other case, when the delays are large enough, there is no synchronization between the network nodes.

\bibliography{plotnikovbib}             

\begin{thebibliography}{23}
\providecommand{\natexlab}[1]{#1}
\providecommand{\url}[1]{\texttt{#1}}
\providecommand{\urlprefix}{URL }
\expandafter\ifx\csname urlstyle\endcsname\relax
  \providecommand{\doi}[1]{doi:\discretionary{}{}{}#1}\else
  \providecommand{\doi}{doi:\discretionary{}{}{}\begingroup
  \urlstyle{rm}\Url}\fi

\bibitem[{Arenas et~al.(2008)Arenas, D\'iaz-Guilera, Kurths, Moreno, and
  Zhou}]{ARE08}
Arenas, A., D\'iaz-Guilera, A., Kurths, J., Moreno, Y., and Zhou, C. (2008).
\newblock Synchronization in complex networks.
\newblock \emph{Phys. Rep.}, 469(3), 93--153.
\newblock \doi{10.1016/j.physrep.2008.09.002}.

\bibitem[{Bryntseva and Fradkov(2019)}]{BRY19}
Bryntseva, T.A. and Fradkov, A.L. (2019).
\newblock Frequency-domain estimates of the sampling interval in multirate
  nonlinear systems by time-delay approach.
\newblock \emph{Intern. J. Control}, 92, 1985--1992.
\newblock \doi{10.1080/00207179.2017.1423394}.

\bibitem[{Churilova(1995)}]{CHU95}
Churilova, M.Y. (1995).
\newblock Analog of the cyclic criterion of absolute stability for systems with
  variable delays.
\newblock \emph{Autom. Remote Control}, 56, 195--198.

\bibitem[{Coombes and Laing(2009)}]{COO09}
Coombes, S. and Laing, C. (2009).
\newblock Delays in activity-based neural networks.
\newblock \emph{Phil. Trans. R. Soc. A}, 367(1891), 1117--1129.
\newblock \doi{10.1098/rsta.2008.0256}.

\bibitem[{Dahlem et~al.(2009)Dahlem, Hiller, Panchuk, and Sch{\"o}ll}]{DAH09}
Dahlem, M.A., Hiller, G., Panchuk, A., and Sch{\"o}ll, E. (2009).
\newblock Dynamics of delay-coupled excitable neural systems.
\newblock \emph{Int.~J.~Bifur.~Chaos}, 19(2), 745--753.
\newblock \doi{10.1142/S0218127409023111}.

\bibitem[{Ermentrout and Ko(2009)}]{ERM09}
Ermentrout, B. and Ko, T.W. (2009).
\newblock Delays and weakly coupled neuronal oscillators.
\newblock \emph{Phil. Trans. R. Soc. A}, 367(1891), 1097--1115.
\newblock \doi{10.1098/rsta.2008.0259}.

\bibitem[{Fradkov(2007)}]{FRA07}
Fradkov, A. (2007).
\newblock \emph{Cybernetical physics: {From} control of chaos to quantum
  control}.
\newblock Springer-Verlag, Berlin Heidelberg.
\newblock \doi{10.1007/978-3-540-46277-4}.

\bibitem[{Geng et~al.(2014)Geng, Zhou, Zhao, Yuan, Ma, and Wang}]{GEN14}
Geng, S., Zhou, W., Zhao, X., Yuan, Q., Ma, Z., and Wang, J. (2014).
\newblock Bifurcation and oscillation in a time-delay neural mass model.
\newblock \emph{Biol. Cybern.}, 108(6), 747--756.
\newblock \doi{10.1007/s00422-014-0616-4}.

\bibitem[{Gorshkov et~al.(2017)Gorshkov, Plotnikov, and Fradkov}]{GOR17}
Gorshkov, A.A., Plotnikov, S.A., and Fradkov, A. (2017).
\newblock Bifurcation and synchronization analysis of neural mass model
  subpopulations.
\newblock \emph{IFAC-PapersOnLine}, 50(1), 14741--14745.
\newblock \doi{10.1016/j.ifacol.2017.08.2577}.

\bibitem[{Grimbert and Faugeras(2006)}]{GRI06}
Grimbert, F. and Faugeras, O. (2006).
\newblock Bifurcation analysis of {Jansen's} neural mass model.
\newblock \emph{Neur. Comput.}, 18(12), 3052--3068.
\newblock \doi{10.1162/neco.2006.18.12.3052}.

\bibitem[{Herbert-Read(2016)}]{HER16}
Herbert-Read, J.E. (2016).
\newblock Understanding how animal groups achieve coordinated movement.
\newblock \emph{J. Exp. Biol.}, 219(19), 2971--–2983.
\newblock \doi{10.1242/jeb.129411}.

\bibitem[{Hong and Strogatz(2011)}]{HON11}
Hong, H. and Strogatz, S.H. (2011).
\newblock Kuramoto model of coupled oscillators with positive and negative
  coupling parameters: An example of conformist and contrarian oscillators.
\newblock \emph{Phys. Rev. Lett.}, 106, 054102.
\newblock \doi{10.1103/PhysRevLett.106.054102}.

\bibitem[{Jansen and Rit(1995)}]{JAN95}
Jansen, B. and Rit, V. (1995).
\newblock Electroencephalogram and visual evoked potential generation in a
  mathematical model of coupled cortical columns.
\newblock \emph{Biol. Cybern.}, 73(4), 357--366.
\newblock \doi{10.1007/BF00199471}.

\bibitem[{Panteley and Lor{\'i}a(2017)}]{PAN17}
Panteley, E. and Lor{\'i}a, A. (2017).
\newblock Synchronization and dynamic consensus of heterogeneous networked
  systems.
\newblock \emph{IEEE Trans. Automat. Control}, 62(8), 3758--3773.
\newblock \doi{10.1109/TAC.2017.2649382}.

\bibitem[{Plotnikov and Fradkov(2018)}]{PLO18}
Plotnikov, S.A. and Fradkov, A.L. (2018).
\newblock On synchronization in fitzhugh-nagumo networks with small delays.
\newblock In \emph{2018 European Control Conference (ECC)}, 2052--2056.
\newblock \doi{10.23919/ECC.2018.8550552}.

\bibitem[{Plotnikov and Fradkov(2021)}]{PLO21}
Plotnikov, S.A. and Fradkov, A.L. (2021).
\newblock Synchronization of nonlinearly coupled networks based on circle
  criterion.
\newblock \emph{Chaos}, 31(10), 103110.
\newblock \doi{10.1063/5.0055814}.

\bibitem[{Proskurnikov(2013)}]{PRO13b}
Proskurnikov, A.V. (2013).
\newblock Average consensus in networks with nonlinearly delayed couplings and
  switching topology.
\newblock \emph{Automatica}, 49(9), 2928--2932.
\newblock \doi{10.1016/j.automatica.2013.06.007}.

\bibitem[{Ren and Beard(2008)}]{REN08}
Ren, W. and Beard, R.W. (2008).
\newblock \emph{Distributed consensus in multi-vehicle cooperative control:
  theory and applications}.
\newblock Springer-Verlag, London.
\newblock \doi{10.1007/978-1-84800-015-5}.

\bibitem[{Schnitzler et~al.(2009)Schnitzler, Munks, Butz, Timmermann, and
  Gross}]{SCH09}
Schnitzler, A., Munks, C., Butz, M., Timmermann, L., and Gross, J. (2009).
\newblock Synchronized brain network associated with essential tremor as
  revealed by magnetoencephalography.
\newblock \emph{Mov. Disorders}, 24(11), 1629--1635.
\newblock \doi{10.1002/mds.22633}.

\bibitem[{Selivanov et~al.(2015)Selivanov, Fradkov, and Fridman}]{SEL15}
Selivanov, A., Fradkov, A., and Fridman, E. (2015).
\newblock Passification-based decentralized adaptive synchronization of
  dynamical networks with time-varying delays.
\newblock \emph{J. Franklin Inst.}, 352(1), 52--72.
\newblock \doi{10.1016/j.jfranklin.2014.10.007}.

\bibitem[{Song et~al.(2009)Song, Makarov, and Velarde}]{SON09}
Song, Y., Makarov, V.A., and Velarde, M.G. (2009).
\newblock Stability switches, oscillatory multistability, and spatio-temporal
  patterns of nonlinear oscillations in recurrently delay coupled neural
  networks.
\newblock \emph{Biol. Cybern.}, 101(2), 147--167.
\newblock \doi{10.1007/s00422-009-0326-5}.

\bibitem[{Steur et~al.(2012)Steur, Oguchi, van Leeuwen, and Nijmeijer}]{STE12}
Steur, E., Oguchi, T., van Leeuwen, C., and Nijmeijer, H. (2012).
\newblock Partial synchronization in diffusively time-delay coupled oscillator
  networks.
\newblock \emph{Chaos}, 22, 043144.
\newblock \doi{10.1063/1.4771665}.

\bibitem[{Sumpter(2010)}]{SUM10}
Sumpter, D.J. (2010).
\newblock \emph{Collective Animal Behavior}.
\newblock Princeton University Press, Princeton.

\end{thebibliography}

\end{document}